# Fault Predictions in Object Oriented Software

Bremananth R and Thushara R

*Abstract*—**The dynamic software development organizations optimize the usage of resources to deliver the products in the specified time with the fulfilled requirements. This requires prevention or repairing of the faults as quick as possible. In this paper an approach for predicting the run-time errors in java is introduced. The paper is concerned with faults due to inheritance and violation of java constraints. The proposed fault prediction model is designed to separate the faulty classes in the field of software testing. Separated faulty classes are classified according to the fault occurring in the specific class. The results are papered by clustering the faults in the class. This model can be used for predicting software reliability.**

## I. INTRODUCTION

The programs written using object-oriented languages may have data flow anomalies and faults. Occasionally one of these faults manifests a failure, and corrective measures are then usually taken to eliminate the fault. The traditional testing techniques that are available primarily focus on the syntactic and semantic error constructs. The other constructs which lead to an incorrect output is due to the fault that occurs during run-time. Our paper takes these types of errors into consideration and errors are predicted during the compilation time, reducing the errors during the run-time.

The errors are taken as metrics to predict fault classes.

The metrics considered are:
1. Spaghetti Error
2. Inconsistent Type Usage error
3. Lvalue required
4. Undefined loop Exception
5. Illicit File Usage Exception
6. Incorrect inheritance

A. *Proposed System*

The preliminary work done by us is to develop a new compiler which is used to test the classes for the above specified metrics and the faulty classes are displayed along with their metrics. The faults are clustered with the common type of errors occurring in it and the output is displayed according to the cluster of the errors that have occurred in the tested classes.

## II METHODOLOGY

The experiment has been carried out using 20 Java classes from different sources and all the classes are measured with 6 metrics. For this, a compiler is generated for the above specified metrices. The compiler compiles the source code and separates the faulty classes containing the errors along with the class name [1].

*A Spaghetti Error*

The error has derived its name from catering science representing a cup of noodles. A cup of noodles remain in an attached format, such that each one is extended with the other. Similarly in multilevel inheritance the derived class is inherited from the base class. Multilevel inheritance is not an error in java but when the inheritance level reaches six or more than six, it leads to a run time anomaly. Figure1 is an example of multilevel inheritance, where A.Java is the super class of the inheritance level for which B.java is a descendant class which in turn is inherited by the class C.java class.

```
Multilevel inheritance (MLI) 1
public class ML_A
{
   ML_A()
   {
    System.out.print("Welcome to ML_A");
   }}
// MLI 2 , MLI 3 , MLI 4 , MLI 5......
class ML_G extends ML_F
{
       ML_G()
       {
       System.out.print("Welcome to ML_G");
       }
       public static void main(String arg[])
       {
          ML_G mlf = new ML_G();
       }}
```

Fig. 1 Illustration of spaghetti error

C.java is base class for the class D.java which serves as the base class of E.java. E.java class is inherited by the class F.java. When the class G.java extends F.java class it leads to an anomaly in the inherited level (as the inheritance level reaches six). Result of spaghetti error is illustrated in figure2.

Bremananth R is Professor with Dept. of Computer Applications, Sri Ramakrishna Engineering College, Coimbatore, Tamil Nadu, India-641008(91-422-2461588,2460088, e-mail: bremresearch@gmail.com).

Thushara R is a research student with Dept. of Computer Technology, Sri Ramakrishna Engineering College, Coimbatore, Tamil Nadu, India.(e-mail: ct.thushara@gmail.com).

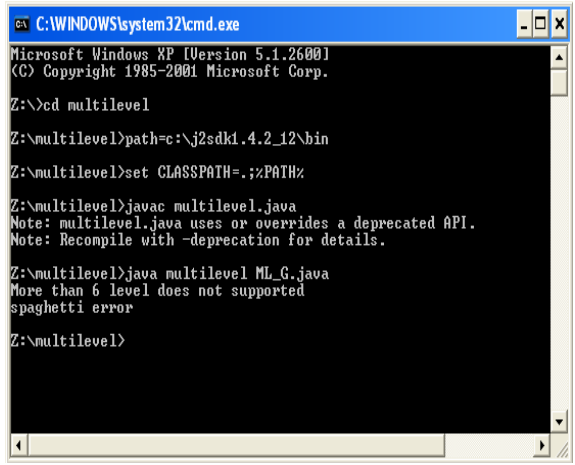

Fig. 2 Output of Spaghetti error

*B Inconsistent Type Usage error (ITU)*

For this fault type [2], a descendant class does not override any inherited method. Thus, there can be no polymorphic behavior. Every instance of a descendant class that is used where an instance of base class is expected can only behave exactly like an instance of base class. That is, only methods of base class can be used. Any additional methods specified in base class are hidden since the instance of derived class is being used as if it is an instance of base class. However, anomalous behavior is still possible. If an instance of descendant class is used in multiple contexts anomalous behavior can occur if derived class has extension methods. In this case, one or more of the extension methods can call a method of base class or directly define a state variable inherited from base class. Anomalous behavior will occur if either of these actions results in an inconsistent inherited state. Figure 3 is an example illustrating the error Inconsistent Type Usage (ITU). In this program stack is a derived class of vector. g() is a function which expects vector as its argument type. But in the above example stack object s is passed where the vector instance v is required. This is syntactically correct because an instance of stack is also an instance of vector.

```
1    public void f (Stack s)
2    {
3        String s1 = "s1";
4        String s2 = "s2";
5        String s3 = "s3";
6        ...
7        s.push (s1);
8        s.push (s2);
9        s.push (s3);
10
11       g (s);
12
13       s.pop ();
14       s.pop ();
15       // Oops!  The stack is empty!
16       s.pop ();
17       ...
18   }
19   public void g (Vector v)
20   {
21       // Remove the last element
22       v.removeElementAt (v.size()-1);
23   }
```

Fig. 3 Illustration of ITU error

The problem begins at line 21 where the last element of s is removed. The fault is manifested when control returns and reaches the first call to *stack::pop ()* at line 14. Here, the element removed from the stack is not the last element that was added, thus the stack integrity constraint will be violated. Hence the result of the software class will be affected. Result of ITU error is illustrated in figure4.

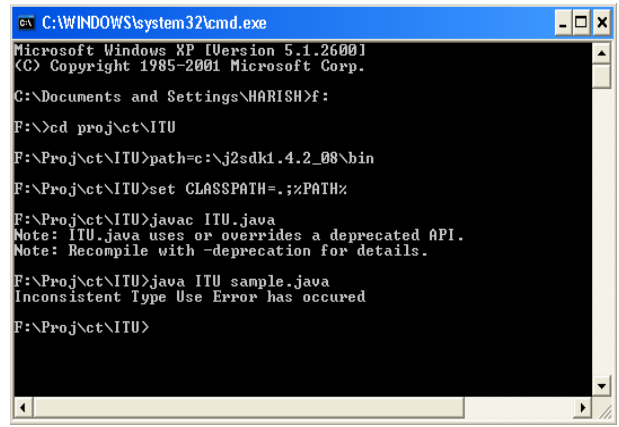

Fig. 4 Output of ITU error

*C Lvalue required*

Comparing of two strings in java can be done by using the .(dot) equals operation. But this condition is violated when the operator == is used to check whether the strings are equal. This is never specified as a fault in java, even though the constraint is violated. Hence our module deals with this constraint specifying that the use of ==operator to equal two strings is illegal. An example program for Lvalue required, given in figure 5 has two string variables d and e. The use of ==operator in the line 8, is against the rules and hence the error message is displayed as Lvalue required.

```
1    class A
2    {
3        int a, b, c, x;
4        String d="WEL";
5        String e="WEL";
6        public void A_a()
7        {
8            if(d==e)
9            {
10           }
11           System.out.println("Class A called");
12       }}
```

Fig. 5 Occurrence of Lvalue required error

Result for Lvalue required is illustrated in figure 6.

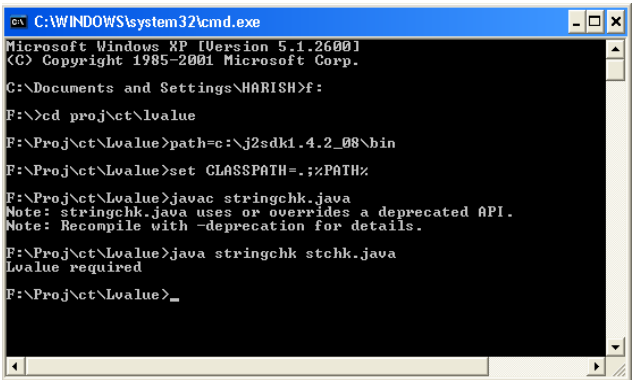

Fig. 6 Output for Lvalue

*D Undefined loop exception*

As the name specifies this error is nothing but a dummy loop. If we use any undefined loop java compiler does not consider this as an error. Our module will consider this as error while at compilation time. In figure7, there is a dummy do while loop which is of no use. Just the loop goes on and on. Hence this is specified as an error by our compiler.

```
1    class A
2    {
3      int a, b, c,x;
4      public void A_a()
5      {
6        if(a>4)
7        {
8          a--;
9        }
10       do
11       {
12       }while(a>10);
13    }}
```

Fig. 7 Illustration of undefined loop exception

Result of undefined loop exception is illustrated in figure 8.

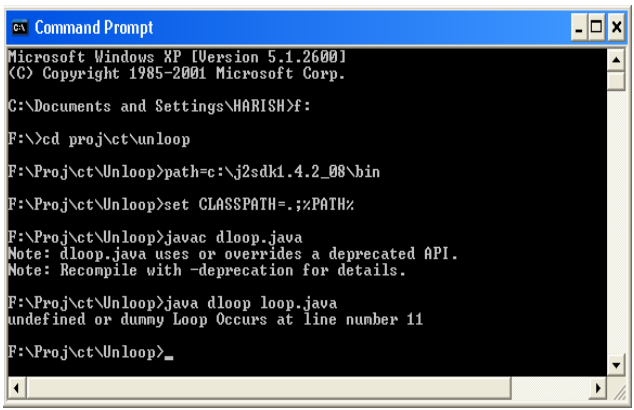

Fig. 8 Output for Undefined loop exception

*E Illicit file usage exception*

Illicit means improper. The proper way of using the files in java is that all the opened files must be closed before it is used again. But java compiler does not specify this error but it will generate some technical error dynamically. Our compiler detects this fault and specifies that the specific file which is opened is not closed. Figure 9 is an example in which two files file _output and data_out opened. But only the file_output file is closed in the line 16 and the file data_out remain unclosed.

```
1 class loopa
2 {
3    loopa ()
4    {
5      int a=0;
6      int i=0;
7      try
8      {
9        FileOutputStream file_output=new FileOutputStream(file);
10       DataOutputStream data_out=new DataOutputStream(file_output);
11       for (i = 0;i < 10;i++)
12       {
13             data_out.writeInt(i);
14             data_out.writeDouble(i);
15       }
16             file_output.close();
17    }
18    catch (IOException e)
19    {
20       System.out.println("IO exception: " + e);
21       }}}
```

Fig. 9 Illustration of illicit file usage exception

The unclosed file leads to a dynamic error during execution when the file is tried to open again. The result of illicit file usage exception is displayed as in figure 10.

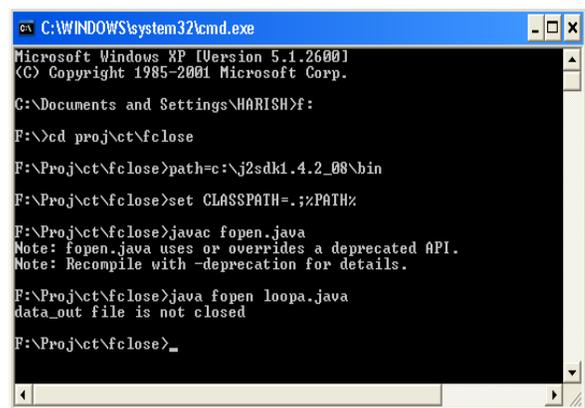

Fig. 10 Output of illicit file usage exception

*F Incorrect Inheritance*

The name itself specifies that the inheritance is not right. The type of inheritance that is incorrect in java is the multiple inheritance, which is replaced with the interfaces. Though java does not support multiple inheritance it does not show a bug that the fault occurring is due to the usage of the incorrect inheritance (i.e.) multiple inheritance. It just gives a syntax error as "{expected". But our module exactly specifies that the error is due to the usage of the incorrect inheritance. Figure11 is an example in which C.Java inherits two classes B.java and A.java. Since a single class inherits two classes, it leads to incorrect inheritance.

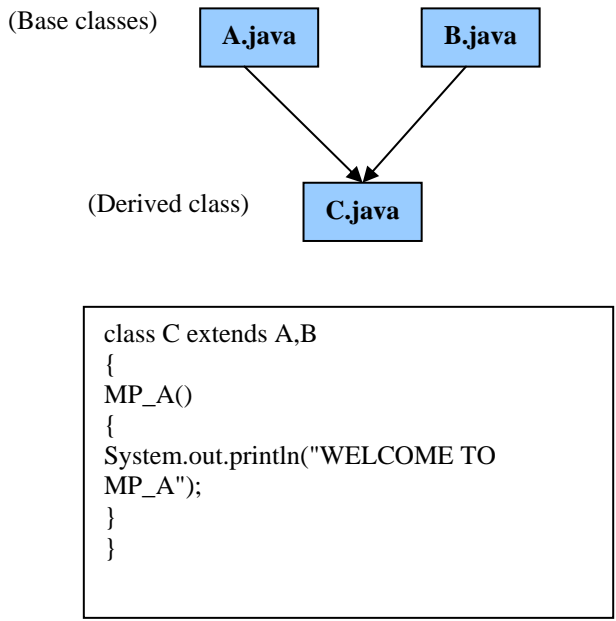

```
class C extends A,B
{
MP_A()
{
System.out.println("WELCOME TO MP_A");
}
}
```

Fig. 11: Illustration of Incorrect Inheritance error

Result of incorrect inheritance error is illustrated in figure12.

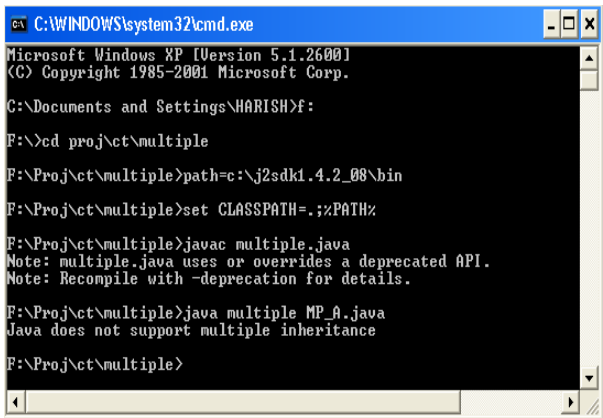

Fig. 12 Output of incorrect inheritance

## IV IMPLEMENTATION AND RESULT ANALYSIS

Our module allows the user to input a folder which contains the classes to be tested. On loading the folder the selected classes to be tested present inside the folder are fed to the source code one by one which monitors for the occurrence of error. The errors are monitored based on the six metrics specified earlier. If an error is detected then the error number is stored in the error list of that particular class in the database. After monitoring all the classes with source code the results are stored in database as in table I.

TABLE I
CLASS NAME WITH THEIR ERROR NUMBERS

| class_name | errlst |
|---|---|
| A.java | 1,6, |
| ML_G.java | 3, |
| ML_H.java | 3,4, |
| MP_A.java | 2,5, |
| loopa.java | 1,6,5, |
| sample.java | 1,4, |

The metrics we have used are stored with the error code and error name as in table II.

TABLE II
ERROR NAME WITH ERROR CODE

| errcode | errname |
|---|---|
| 1 | Lvalue required |
| 2 | Incorrect inheritance error |
| 3 | Spaghetti error |
| 4 | Inconsistent Type Usage error |
| 5 | Illicit file usage exception |
| 6 | Undefined loop exception |

When the compiler has identified the faulty classes the result will be displayed in a format as that of figure 13. The user can view the errors of a particular class by selecting the class required and on clicking the "error view" button the errors of that class will be listed at the right side by retrieving its error names from the database with the help of "errcode "listed in errlst.

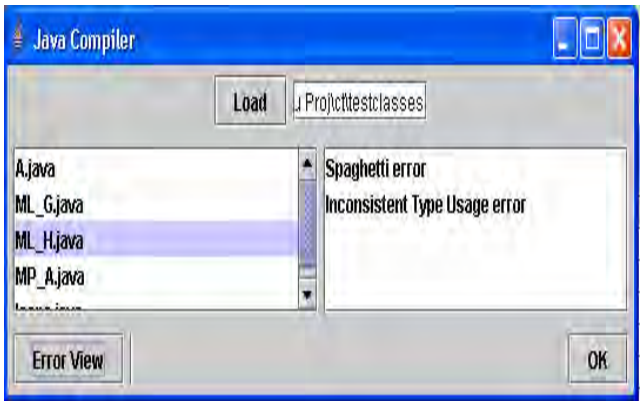

Fig. 13 Error view for a specific class

On clicking the "ok" button all the errors with their specific classes are displayed as in figure 14.

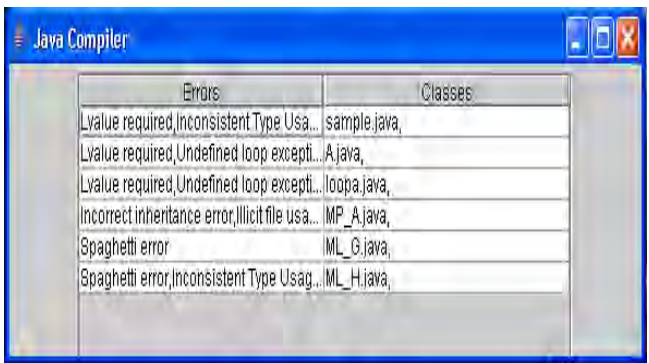

Fig. 14 Clustered errors in various classes

Listing.1 specifies source code which creates up a panel with error view, ok and load buttons. The load button loads the path of the folder where the classes to be tested are placed. The code for compiling the java classes and displaying the errors is called in turn and the classes with the errors are displayed as in figure 13.

```
public resultwin() {
    super(new BorderLayout());
    fileName = new JTextField(10);
    errlst = new Vector();
    Load = new JButton("Load");
    toppanel = new JPanel();
    toppanel.add(Load);
    toppanel.add(fileName);
     //adding List controldetails
   Load.addActionListener(new LoadListener());
    listModel = new DefaultListModel();
    listModel1 = new DefaultListModel();
//Create the list and put it in a scroll pane.
     list = new JList(listModel);
     list1 = new JList(listModel1);
list.setSelectionMode(ListSelectionModel.SINGLE_SELECTION);
   list.setSelectedIndex(0);
     list.addListSelectionListener(this); //list control event handled
     list.setVisibleRowCount(5);
     JScrollPane listScrollPane = new JScrollPane(list); //scrollpane set to list control
     list1.setVisibleRowCount(5);
 JScrollPane listScrollPane1 = new JScrollPane(list1);
     JButton hireButton = new JButton(hireString);
     HireListener hireListener = new HireListener(hireButton);
     hireButton.setActionCommand(hireString);
     hireButton.addActionListener(hireListener);
     hireButton.setEnabled(true);
        OK = new JButton("OK");
        OK.addActionListener(new OKListener());
     fireButton = new JButton(fireString);
     fireButton.setActionCommand(fireString);
    fireButton.addActionListener(new FireListener());
//Create a panel that uses BoxLayout.
     JPanel buttonPane = new JPanel();
        JPanel listpanel = new JPanel();
        listpanel.add(listScrollPane);
        listpanel.add(listScrollPane1);
    buttonPane.setLayout(new BoxLayout(buttonPane,
                    BoxLayout.LINE_AXIS));
     buttonPane.add(fireButton);
     buttonPane.add(Box.createHorizontalStrut(5));
      buttonPane.add(new JSeparator(SwingConstants.VERTICAL));
     buttonPane.add(Box.createHorizontalStrut(5));
buttonPane.setBorder(BorderFactory.createEmptyBorder(5,5,5,5));
      add(toppanel, BorderLayout.NORTH);
      add(listpanel, BorderLayout.CENTER);
      add(buttonPane, BorderLayout.PAGE_END);
}
class OKListener implements ActionListener
    {
public void actionPerformed(ActionEvent e)
    {
//adding selected fields to the vector
    Vector v = new Vector();
 v.addElement(list1.getModel().getElementAt(i));
dispwin.createAndShowGUI(); //next screen with vector as input
    }
}
class LoadListener implements ActionListener
    {
public void actionPerformed(ActionEvent e)
    {
     JFileChooser jfc = new JFileChooser();
jfc.setFileSelectionMode(JFileChooser.DIRECTORIES_ONLY);
```

```java
            if (e.getSource() == Load)
              {
int rVal = jfc.showOpenDialog(new resultwin());
if (rVal == JFileChooser.APPROVE_OPTION)
{
    file = jfc.getSelectedFile();
    String fname1 = file.getAbsolutePath();
    fileName.setText(fname1);
    mainfile.disp(fname1);
}
  try
   {
Class.forName("sun.jdbc.odbc.JdbcOdbcDriver");
con =
DriverManager.getConnection("jdbc:odbc:stst");
st = con.createStatement();
r = st.executeQuery("select * from clserr");
while (r.next())
{
listModel.addElement(r.getString("class_name"));
errlst.addElement(r.getString("errlst"));
}
}
catch (Exception eee)
{
}
}
}
}
class FireListener implements ActionListener {
  public void actionPerformed(ActionEvent e) {
  //This method can be called only if there's a valid
//selection so go ahead and remove whatever's
//selected.
    int index = list.getSelectedIndex();
String s = (String)errlst.elementAt(index);
String res[] = p.split(s);
listModel1.clear();
try
{
Class.forName("sun.jdbc.odbc.JdbcOdbcDriver");
Connection con1 =
DriverManager.getConnection("jdbc:odbc:stst");
Statement st1 = con.createStatement();
ResultSet r1;
for (int rr = 0; rr < res.length; rr++)
{
  System.out.println(res[rr].trim());
r1 = st1.executeQuery("select errname from errtab
where errcode='" + res[rr].trim() +"'");
if (r1.next())
{
listModel1.addElement(r1.getString("errname"));
}
}
}
catch (Exception eee)
{
  System.out.print(eee);
}

  int size = listModel.getSize();
  if (size == 0) { //Nobody's left, disable firing.
   fireButton.setEnabled(false);
  }
else { //Select an index.
   if (index == listModel.getSize()) {
      //removed item in last position
      index--;
    }
  list.setSelectedIndex(index);
    list.ensureIndexIsVisible(index);
  }
 }
}
  //This listener is shared by the text field and the hire
button.
    class HireListener implements ActionListener,
DocumentListener {
       private boolean alreadyEnabled = false;
private JButton button;
       public HireListener(JButton button) {
          this.button = button;
       }
       //Required by ActionListener.
       public void actionPerformed(ActionEvent e) {
int index = list1.getSelectedIndex(); //get selected index
String s = (list1.getSelectedValue()).toString();
listModel1.remove(index);
listModel.addElement(s);
 if (index == -1) { //no selection, so insert at beginning
      index = 0;
   } else {         //add after the selected item
         index++;
        }
list1.setSelectedIndex(index);
       list1.ensureIndexIsVisible(index);
     }
protected boolean alreadyInList(String name) {
       return listModel.contains(name);
     }
     //Required by DocumentListener.
   public void insertUpdate(DocumentEvent e) {
   enableButton();
 }
 //Required by DocumentListener.
 public void removeUpdate(DocumentEvent e) {
   handleEmptyTextField(e);
}
//Required by DocumentListener.
public void changedUpdate(DocumentEvent e) {
   if (!handleEmptyTextField(e)) {
   enableButton();
      }
    }
 private void enableButton() {
       if (!alreadyEnabled) {
          button.setEnabled(true);
       }
     }
```

```
private boolean
handleEmptyTextField(DocumentEvent e) {
    if (e.getDocument().getLength() <= 0) {
          button.setEnabled(false);
          alreadyEnabled = false;
          return true;
     }
        return false;
    }  }
//This method is required by
ListSelectionListener.
   public void
valueChanged(ListSelectionEvent e) {
     if (e.getValueIsAdjusting() == false) {

        if (list.getSelectedIndex() == -1) {
        //No selection, disable fire button.
           fireButton.setEnabled(false);
        } else {
        //Selection, enable the fire button.
           fireButton.setEnabled(true);
        }
     }  }
   /**
* Create the GUI and show it.  For thread safety,
    * this method should be invoked from the
    * event-dispatching thread.
    */
   private static void createAndShowGUI() {
       //Create and set up the window.
       JFrame frame = new JFrame("Java Compiler");

frame.setDefaultCloseOperation(JFrame.EXIT_ON_CLOSE);

       //Create and set up the content pane.
       JComponent newContentPane = new resultwin();
       newContentPane.setOpaque(true);
//content panes must be opaque
       frame.setContentPane(newContentPane);

       //Display the window.
       frame.pack();
       frame.setVisible(true);
   }   public static void main(String[] args) {
     //Schedule a job for the event-dispatching thread:
       //creating and showing this application's GUI.
javax.swing.SwingUtilities.invokeLater(new Runnable() {
         public void run() {
            createAndShowGUI();} }); }}
```

Listing. 1 Displaying errors of specific class

The source code compiles the classes to be tested and specifies the faults that occur in a single combination is clustered and the classes that posses those clustered faults are displayed along with the clustered faults(Listing 2)

```
         dispwin()
         {
                Pattern p =
Pattern.compile(",");
              try
              {
Class.forName("sun.jdbc.odbc.JdbcOdbcDriver");
Connection con =
DriverManager.getConnection("jdbc:odbc:stst");
st = con.createStatement();
st1 = con.createStatement();
st2 = con.createStatement();
v = new Vector();
v1 = new Vector();
v2 = new Vector();
r = st.executeQuery("select distinct errlst from
clserr");
while (r.next())
{
v.addElement(r.getString("errlst"));
}
for (int vv = 0; vv < v.size(); vv++)
{
  String dstr = (String)v.elementAt(vv);
  String ress[] = p.split(dstr);
  String ename="";
  for (int i = 0; i < ress.length; i++)
  {
  System.out.println("select errname from errtab
where errcode='" + ress[i].trim() + "'");
   r1 = st1.executeQuery("select errname from
errtab where errcode='" + ress[i].trim() +"'");
    while (r1.next())
   {
     ename += r1.getString("errname") + ",";
      System.out.println(ename);
   }
}
v1.addElement(ename.substring(0,
ename.length() - 1));
System.out.println("select class_name from clserr
where errlst='" + dstr.trim() + "'");
r2 = st2.executeQuery("select class_name from
clserr where errlst=' " + dstr.trim() + "'");
String dstr1 = "";
while (r2.next())
{
dstr1 += r2.getString("class_name")+",";
System.out.println(dstr1);
}
v2.addElement(dstr1);
}
Vector columnNames = new Vector();
columnNames.addElement("Errors");
columnNames.addElement("Classes");
int columns = columnNames.size();
```

```
for (int rrr = 0; rrr < v2.size(); rrr++)
{
   System.out.println(v2.elementAt(rrr));
   Vector row = new Vector(columns);
   row.addElement(v1.elementAt(rrr));
   row.addElement(v2.elementAt(rrr));
   data.addElement(row);
}
JTable table = new JTable(data, columnNames);
JScrollPane scrollPane = new JScrollPane(table);
add(scrollPane, BorderLayout.CENTER);
}
catch (Exception e)
{
   System.out.println(e);
}
}
public static void createAndShowGUI()
{
//Create and set up the window.
JFrame frame = new JFrame("Java Compiler");
frame.setDefaultCloseOperation(JFrame.EXIT_ON_CLOSE);
//Create and set up the content pane.
JComponent newContentPane = new dispwin();
newContentPane.setOpaque(true); //content panes must be opaque
frame.setContentPane(newContentPane);
//Display the window.
frame.pack();
frame.setVisible(true);
}}
```

Listing. 2 Displaying clustered errors with their classes

## V CONCLUSION AND FUTURE ENHANCEMENT

The run-time errors are predicted which would lead to higher efficiency of software and quality of resulting process. The common combination of errors can be identified. By considering the object-oriented faults that can be generated from object-oriented constructs, we can gain insight into a number of issues in object-oriented analysis.

In future, this paper can be used to predict the percentage of failure in object-oriented software due to these faults. The paper can also be enhanced using neural network techniques by increasing the number of metrices which provides high accuracy in discrimination between faulty and fault-free classes.

ACKNOWLEDGMENT

Authors thank the SNR & Sons Charitable Trust, Coimbatore, India for providing necessary infrastructure to complete this research study.

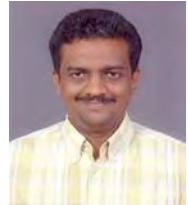

**Bremananth R** received the B.Sc and M.Sc. degrees in Computer Science from Madurai kamaraj and Bharathidsan University, India in 1991 and 1993, respectively. He has obtained M.Phil. degree in Computer Science & Engineering from Bharathiar University. He has received his Ph.D. degree in the Department of CSE, PSG College of Technology, India, Anna University, Chennai.

Presently, He is a Professor Department of Computer Applications, Sri Ramakrishna Engineering College, Coimbatore, India. He has 16 years of teaching experience and published several research papers in the National and International Journals and Conferences. He has received M N Saha Memorial award for the year 2006 by IETE. His fields of research are pattern recognition, computer vision, image processing, biometrics, multimedia and soft computing.

Dr. Bremananth is a member of Indian society of technical education, advanced computing society, ACS and IETE.

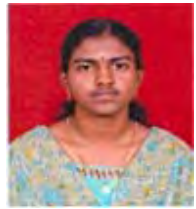

**Thushara R** has completed her B.Sc. in Computer Technology in Sri Ramakrishna Engineering College, affiliated to Anna University Coimbatore, Tamil Nadu, India. She is also a research student with Dept. of Computer Applications Sri Ramakrishna Engineering College, affiliated to Anna University Coimbatore, Tamil Nadu, India.